\documentclass{jpsj3}
\usepackage{txfonts}

\title{Phase Separation Induced by Symmetric Monocycle Optical Pulse in Extended Hubbard Models}

\author{Hiroki Yanagiya, Yasuhiro Tanaka, and Kenji Yonemitsu\thanks{E-mail: kxy@phys.chuo-u.ac.jp}}
\inst{Department of Physics, Chuo University, Bunkyo, Tokyo 112-8551, Japan} %\\

\abst{ Many-electron dynamics induced by a symmetric monocycle electric-field pulse of large amplitude is theoretically investigated in one- and two-dimensional half-filled extended Hubbard models on regular lattices (i.e., without dimerization) using the exact diagonalization method for small systems and the Hartree-Fock approximation for large systems. The formation of a negative-temperature state and the change from repulsive interactions to effective attractive interactions are shown to be realized for a wide region of the field amplitude and the excitation energy. For a nonnegligible intersite repulsive interaction, the numerical results are consistent with the fact that the phase separation between charge-rich and charge-poor regions is caused by the corresponding effective attraction. }

\begin{document}
\maketitle

\section{Introduction}
Physical phenomena caused by intense optical pulses have attracted much attention. For relatively weak pulses, various photoinduced phase transitions are known to take place in solids.\cite{koshigono_jpsj06,yonemitsu_pr08} Here, however, something qualitatively different is expected. Among the states that have been proposed for intense pulses, we focus on negative-temperature states.\cite{tsuji_prl11,tsuji_prb12} For systems whose energy spectra are bounded above, negative-temperature states are realized when high-energy levels are occupied in such a way that the energy of the system is higher than that of an equilibrium state at infinite temperature.\cite{purcell_pr51,ramsey_pr56,klein_pr56} To form such a state, we need to increase the energy of the system (i.e., occupy high-energy levels selectively), suppressing the rise in the entropy (i.e., not merely occupying various energy levels). 

Two-level systems are typical ones for which negative-temperature states are discussed. Many molecular materials have dimerized structures in which bonding and antibonding orbitals exist. They can be regarded as interacting two-level systems that are linked through interdimer electron transfers. The quasi-two-dimensional metal complex Et$_2$Me$_2$Sb[Pd(dmit)$_2$]$_2$ (dmit = 1,3-dithiol-2-thione-4,5-dithiolate) is one of those strongly dimerized molecular materials for which photoinduced phase transitions have been experimentally\cite{ishikawa_prb09} and theoretically\cite{nishioka_jpsj13a,nishioka_jpsj13b,nishioka_jpsj14} studied. It has a charge-ordered ground state in which charge-rich and charge-poor dimers are regularly arrayed. The model used for this material is greatly simplified, keeping the dimerized structure, into a one-dimensional three-quarter-filled strongly dimerized extended Hubbard model with electron-phonon couplings, for which the formation of a negative-temperature state and the change from repulsive interactions to effective attractive interactions are demonstrated in the band-insulator phase using the exact diagonalization method.\cite{yonemitsu_jpsj15} This model is close to interacting two-level systems in that the intradimer transfer integral is much larger than the interdimer transfer integral. 

For regular systems without dimerization, Tsuji et al. have shown such changes of interactions in negative-temperature states for continuous waves,\cite{tsuji_prl11} half-cycle pulses, and asymmetric monocycle pulses\cite{tsuji_prb12} using the dynamical mean-field theory for the half-filled Hubbard model. For continuous waves, the modulation of the effective transfer integral is known in the context of dynamical localization\cite{dunlap_prb86,grossmann_prl91,kayanuma_pra94} and given by the time average of the transfer integral with the Peierls phase factor. For very short pulses, the modulation is explained through the dynamical phase shift on the basis of the sudden approximation.\cite{tsuji_prb12} 

The dynamical phase shift is proportional to the time integral of the electric field. In the sudden approximation, when the dynamical phase shift is $ \pi $, the sign of the Peierls phase factor is instantaneously inverted and the band structure is consequently inverted. For symmetric monocycle pulses, the time integral of the electric field is zero, so that the dynamical phase shift vanishes. For finite pulse widths, symmetric monocycle pulses make little contribution, and their effects are not sufficiently large to produce a negative-temperature state, in the dynamical mean-field theory as described in Ref.~\citen{tsuji_prb12}. However, negative-temperature states are realized by symmetric monocycle pulses, as described in Ref.~\citen{yonemitsu_jpsj15}, and explained by the total-energy increments.\cite{nishioka_jpsj14} The modulation of the effective transfer integral is defined through the total-energy increment and described by the zeroth-order Bessel function, as in the dynamical localization for continuous waves, in the two-level system.\cite{nishioka_jpsj14} 

As mentioned above, the strongly dimerized model used in Ref.~\citen{yonemitsu_jpsj15} is close to interacting two-level systems. This closeness may appear to be advantageous for the system to increase the energy suppressing the rise in the entropy. Regarding the fact that the repulsion-to-attraction conversion is hardly realized by symmetric monocycle pulses applied to the half-filled Hubbard model, in the dynamical mean-field theory as described in Ref.~\citen{tsuji_prb12}, is the distance of the regular Hubbard model from interacting two-level systems the reason? In this paper, we show that the answer is no: the formation of a negative-temperature state and the change from repulsive interactions to effective attractive interactions are achieved by a symmetric monocycle pulse applied to the one- and two-dimensional half-filled extended Hubbard models on regular lattices, i.e., without dimerization. However, the total-energy increment is no longer described by the zeroth-order Bessel function. 

Photoinduced superconductivity is discussed in Ref.~\citen{tsuji_prl11} using the Hubbard model that has only the on-site interaction $ U $. Real substances always have intersite interactions, which are important for the exciton effect. From the viewpoint of the possible inversion of interactions, the intersite interactions are important because they can cause phase separation once they are converted into effective attractive interactions. In this paper, we consider extended Hubbard models, which have the nearest-neighbor interaction $ V $ in addition. 

For phase separation, many theoretical studies have been performed using the one-dimensional\cite{ogata_prl91} and two-dimensional\cite{hellberg_prl97} $t$-$J$ models as well as the one-dimensional,\cite{voit_prb92,penc_prb94,kuroki_prb94,clay_prb99,aligia_prb00} two-dimensional,\cite{su_prb04} and even higher-dimensional\cite{vandongen_prl95} extended Hubbard models at\cite{voit_prb92} and near\cite{vandongen_prl95}  half filling, at quarter filling,\cite{penc_prb94} and at general fillings.\cite{kuroki_prb94,clay_prb99,aligia_prb00,su_prb04} Many of them have been motivated by a desire to clarify the mechanism of high-temperature superconductivity. A superconducting state is indeed anticipated near the instability toward phase separation. It is now known that phase separation is realized in a rather wide region in the plane spanned by the interaction strength and the filling. In general, when the intersite interaction is attractive, phase separation is easily achieved.\cite{voit_prb92,penc_prb94,kuroki_prb94,su_prb04} Thus, when the system has an intersite interaction and the interactions are expected to be inverted by an intense pulse, one must consider the possibility for phase separation. 
In this paper, we employ the exact diagonalization method for small systems and the Hartree-Fock approximation for large systems to show that phase separation is indeed possibly induced by an intense symmetric monocycle optical pulse. 

\section{Extended Hubbard Model on Regular Lattice}

\begin{figure}
\includegraphics[height=9cm]{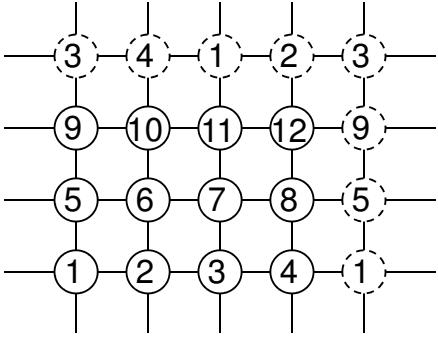}
\caption{%(Color online) 
$ 4 \times 3 $ lattice with periodic boundary condition.
\label{fig:lattice_str}}
\end{figure}

Many-electron dynamics after strong photoexcitation is studied in the half-filled extended Hubbard model, 
\begin{equation}
H = -t_0 \sum_{\langle i,j \rangle, \sigma} 
\left(c^\dagger_{i,\sigma} c_{j,\sigma}+ c^\dagger_{j,\sigma} c_{i,\sigma} \right)
%\nonumber \\ & & 
+U \sum_i \left( n_{i,\uparrow}-\frac12 \right)
\left( n_{i,\downarrow}-\frac12 \right) 
+V \sum_{\langle i,j \rangle} ( n_i-1 ) ( n_j-1 ) 
\;, \label{eq:hamiltonian}
\end{equation}
where $ c_{i,\sigma}^\dagger $ creates an electron with spin $ \sigma $ at site $ i $, $ n_{i,\sigma} = c_{i,\sigma}^\dagger c_{i,\sigma} $, and $ n_i = \sum_\sigma n_{i,\sigma} $. The transfer integral is denoted by $ t_0 $. The parameter $ U $ represents the on-site repulsion strength, and $ V $ represents the nearest-neighbor repulsion strength. For the one-dimensional lattice, we use the 12-site periodic chain. For the two-dimensional lattice, we use the $ 4 \times 3 $-site system with the periodic boundary condition shown in Fig.~\ref{fig:lattice_str}. 

Before the photoexcitation, the system is in the ground state obtained by the exact diagonalization method unless stated otherwise. Photoexcitation is introduced through the Peierls phase 
\begin{equation}
c_{i,\sigma}^\dagger c_{j,\sigma} \rightarrow
\exp \left[
\frac{ie}{\hbar c} \mbox{\boldmath $r$}_{ij} \cdot \mbox{\boldmath $A$}(t)
\right] c_{i,\sigma}^\dagger c_{j,\sigma}
\;, \label{eq:photo_excitation}
\end{equation}
where $ \mbox{\boldmath $r$}_{ij}=\mbox{\boldmath $r$}_j-\mbox{\boldmath $r$}_i $ and $ \mbox{\boldmath $r$}_i $ being the location of the $ i $th site. The right-hand side is substituted into Eq.~(\ref{eq:hamiltonian}). For symmetric monocycle electric-field pulses, we use the time-dependent vector potential 
\begin{equation}
\mbox{\boldmath $A$} (t) = \frac{c\mbox{\boldmath $F$}}{\omega} \left[ \cos (\omega t)-1 \right] 
\theta (t) \theta \left( \frac{2\pi}{\omega}-t \right)
\;, \label{eq:monocycle_pulse}
\end{equation}
where $ \mbox{\boldmath $F$}=(F,F) $ is polarized in the (1,1) direction on the square lattice ($ \mbox{\boldmath $F$}=F $ on the chain) and the central frequency $ \omega $ is chosen to be nearly resonant with the optical gap unless stated otherwise. 
The time-dependent Schr\"odinger equation is numerically solved by expanding the exponential evolution operator with a time slice $ dt $=0.01 to the 15th order and by checking the conservation of the norm.\cite{yonemitsu_prb09} 

The kinetic energy is defined as the expectation value of the first term in Eq.~(\ref{eq:hamiltonian}). The total energy is the expectation value of Eq.~(\ref{eq:hamiltonian}), which becomes zero in equilibrium at infinite temperature. We set $ t_0 $=1 as a unit of energy. The interaction parameters $ U $ and $ V $ are varied. The intersite distance is set to be $ a $ in any direction. The time-averaged double occupancy is calculated by 
\begin{equation}
\langle \langle n_{i,\uparrow} n_{i,\downarrow} \rangle \rangle =
\frac{1}{t_w} \int_{t_s}^{t_s+t_w} 
\langle \Psi (t) \mid n_{i,\uparrow} n_{i,\downarrow} \mid \Psi (t) \rangle dt
\;, \label{eq:time_average}
\end{equation}
with $ t_s=5 T $, $ t_w=45 T $, and $ T=2\pi/\omega $ being the period of the oscillating electric field. The other time-averaged quantities are calculated likewise. 

The systems to which the exact diagonalization method can be applied are quite small, and their nonequilibrium dynamics basically from the ground state can be discussed. On the other hand, the dynamical mean-field theory can be used to discuss the thermodynamic limit at finite temperatures from the beginning, but it uses a mapping to a single-impurity problem, so that events in the momentum space such as Fermi surface nesting or the Umklapp process are included indirectly through the self-energy that is in the second order with respect to electron transfers. Thus, it often underestimates the tendency toward an insulator. Here only, we use a half-cycle pulse and compare numerical results obtained by the present exact diagonalization method with those in the dynamical mean-field theory for the half-filled Hubbard model ($ V = 0 $).\cite{tsuji_prb12} The details are shown in the Appendix. In both approaches, we find (i) that negative-temperature states appear when the Peierls phase shift after the photoexcitation is near an odd multiple of $ \pi $ and $ U/(4t_0) $ is not too large, and (ii) that, as the pulse width increases and $ U/(4t_0) $ increases, the region where negative-temperature states appear is shifted toward a larger Peierls phase shift. These findings suggest that, for the formation of a negative-temperature state, it does not matter whether the initial state is metallic as in Ref.~\citen{tsuji_prb12} or insulating as in the present study for $ U \neq 0 $. 

\section{States after Symmetric Monocycle Pulse Excitation in Hubbard Model}

\begin{figure}
\includegraphics[height=11.2cm]{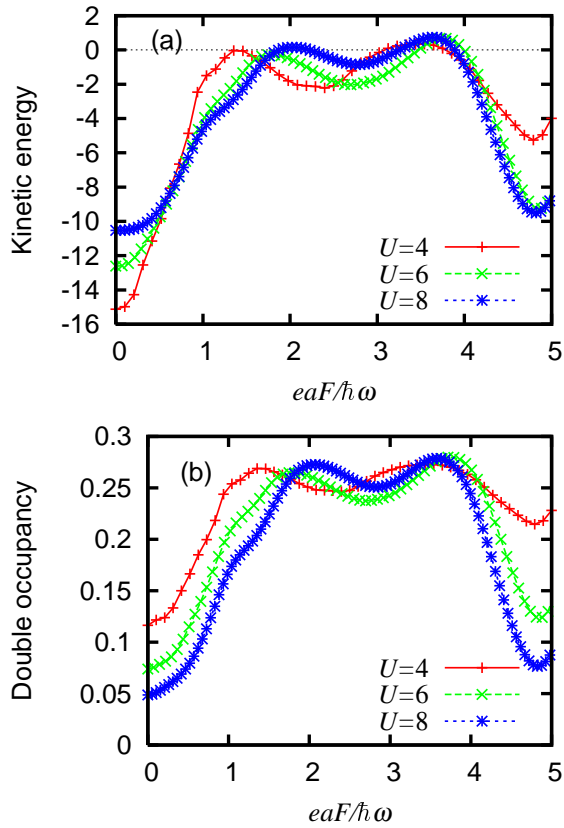}
\caption{(Color online) 
(a) Time-averaged kinetic energy and (b) time-averaged double occupancy $ \langle \langle n_{i,\uparrow} n_{i,\downarrow} \rangle \rangle $ as functions of $ eaF/\hbar\omega $ for different values of $ U $. The other parameters are $ V=0 $, $ \omega=2.4 $ for $ U=4 $, $ \omega=3.43 $ for $ U=6 $, and $ \omega=4.82 $ for $ U=8 $. 
\label{fig:4x3_Ux_w1p0EoptNS_fx_0to5}}
\end{figure}

Because the results for the one- and two-dimensional cases are similar, we show them in the two-dimensional case. Time-averaged quantities are shown in Fig.~\ref{fig:4x3_Ux_w1p0EoptNS_fx_0to5} as functions of $ eaF/\hbar\omega $ for $ U $=4, 6, and 8. 
Around $ eaF/\hbar\omega $=3.7, the time-averaged kinetic energy is positive [Fig.~\ref{fig:4x3_Ux_w1p0EoptNS_fx_0to5}(a)]. This indicates that the momentum distributions at $ -2t_0[\cos(k_x)+\cos(k_y)] > 0 $ are larger than those at $ -2t_0[\cos(k_x)+\cos(k_y)] < 0 $. Note that a symmetric monocycle pulse is used and consequently the Peierls phase returns to the initial value. This is not caused by the band-structure inversion through the dynamical phase shift; Umklapp scattering is considered to be responsible. This change in the momentum distributions is accompanied by a large increase in the total energy. The total energy here is the sum of the kinetic energy and $ U $ times the double occupancy relative to its noninteracting value of 0.25 at half filling. 

The time-averaged double occupancy $ \langle \langle n_{i,\uparrow} n_{i,\downarrow} \rangle \rangle $ indeed becomes larger than 0.25 for an $ eaF/\hbar\omega $ region including the region for positive time-averaged kinetic energies around $ eaF/\hbar\omega $=3.7 [Fig.~\ref{fig:4x3_Ux_w1p0EoptNS_fx_0to5}(b)]. It appears as if these nonequilibrium states possess an attractive on-site interaction. This is caused by the fact that the energy supplied by photoexcitation is distributed to both the kinetic energy and the interaction energy. When the kinetic energy is positive and $ \langle \langle n_{i,\uparrow} n_{i,\downarrow} \rangle \rangle > 0.25 $, the total energy is positive, i.e., it is larger than the value taken by an equilibrium state at infinite temperature. This total energy is treated as an order parameter in Ref.~\citen{tsuji_prb12}. 

In the band-insulator phase of the one-dimensional three-quarter-filled strongly dimerized extended Peierls-Hubbard model, the region of $ eaF/\hbar\omega $ where negative-temperature states appear is almost independent of $ U $, and $ \langle \langle n_{i,\uparrow} n_{i,\downarrow} \rangle \rangle $ in the negative-temperature state (i.e., its effective on-site attraction) increases with $ U $.\cite{yonemitsu_jpsj15} Here, however, the situation is not so simple. The value of $ eaF/\hbar\omega $ where the local maximum of the time-averaged kinetic energy appears or where that of $ \langle \langle n_{i,\uparrow} n_{i,\downarrow} \rangle \rangle $ appears depends on $ U $. Furthermore, in the band-insulator phase of the model above, the energy flow into the electron-electron interaction terms is quite small.\cite{yonemitsu_jpsj15} In the present case, the insulating ground state is caused by the electron-electron interaction term so that the energy flow into this term is relatively large. This is quantitatively discussed later in Sect.~\ref{sec:extended}. 

\begin{figure}
\includegraphics[height=11.2cm]{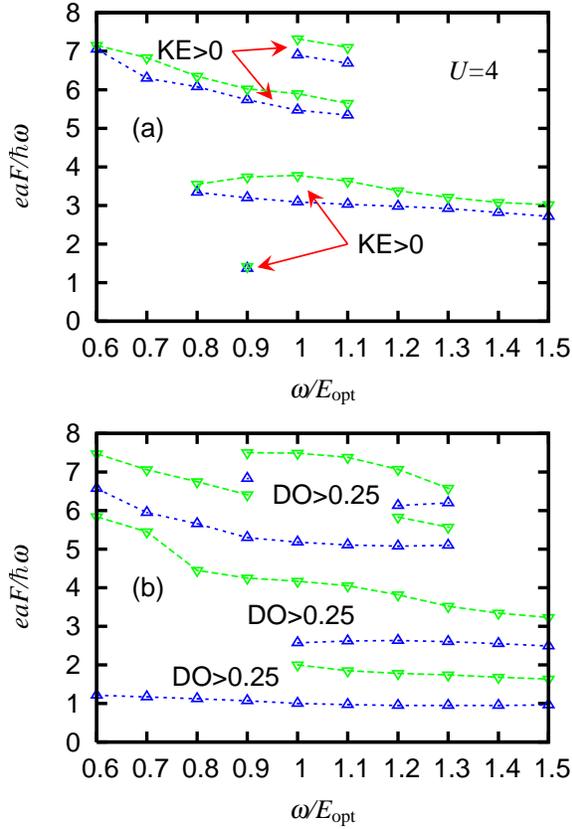}
\caption{(Color online) 
(a) Regions where time-averaged kinetic energy is positive, denoted by ``KE$>$0'', and (b) regions where $ \langle \langle n_{i,\uparrow} n_{i,\downarrow} \rangle \rangle $ is larger than 0.25, denoted by ``DO$>$0.25'', for $ U $=4 in plane spanned by $ eaF/\hbar\omega $ and $ \omega / E_{\rm opt} $ with $ E_{\rm opt} $ being the optical gap.
\label{fig:4x3_U4wxfy_KEDO}}
\end{figure}

Such negative-temperature states accompanied by inverted on-site interactions appear in a wide $ \omega $ region. Their appearance is not limited to the case where $ \omega $ is nearly resonant with the optical gap. For $ U $=4, the regions where the time-averaged kinetic energy is positive and the regions where $ \langle \langle n_{i,\uparrow} n_{i,\downarrow} \rangle \rangle $ is larger than 0.25 are shown in Figs.~\ref{fig:4x3_U4wxfy_KEDO}(a) and \ref{fig:4x3_U4wxfy_KEDO}(b), respectively. 
Here, the plane is spanned by $ eaF/\hbar\omega $ and $ \omega / E_{\rm opt} $ with $ E_{\rm opt} $ being the optical gap. For each $ \omega / E_{\rm opt} $, the upper and lower bounds of the region(s) are denoted by inverted triangles and triangles, respectively. For $ \omega $=0.6$ E_{\rm opt} $ and 0.7$ E_{\rm opt} $, $ eaF/\hbar\omega $ must be larger than 6 for the time-averaged kinetic energy to be positive, but $ \langle \langle n_{i,\uparrow} n_{i,\downarrow} \rangle \rangle > 0.25 $ is widely achieved. For $ \omega $=1.4$ E_{\rm opt} $ and 1.5$ E_{\rm opt} $, $ eaF/\hbar\omega $ must be less than 4 for a positive time-averaged kinetic energy and for $ \langle \langle n_{i,\uparrow} n_{i,\downarrow} \rangle \rangle > 0.25 $. In this sense, the resonant condition is optimum for these two phenomena and consequently for a negative-temperature state to be formed. The regions of $ eaF/\hbar\omega $ for $ \langle \langle n_{i,\uparrow} n_{i,\downarrow} \rangle \rangle > 0.25 $ contain those for positive time-averaged kinetic energies. Therefore, the former regions are wider than the latter. In the one-dimensional case, negative-temperature states accompanied by inverted on-site interactions also appear in a wide $ U $,$ \omega $ region. 

\section{States after Symmetric Monocycle Pulse Excitation in Extended Hubbard Model \label{sec:extended}}

If both the on-site and intersite interactions are inverted, we expect that the effective attractive interactions will cause phase separation into a region of high-charge-density sites and a region of low-charge-density sites, although their boundaries fluctuate quantum-mechanically. How easily the phase separation is realized depends on the dimensionality through the nature of the fluctuating boundary. Thus, below we compare the one-dimensional case, where the boundary consists of two points, with the two-dimensional case, where the boundary consists of a line. 

\subsection{One-dimensional case}

\begin{figure}
\includegraphics[height=5.6cm]{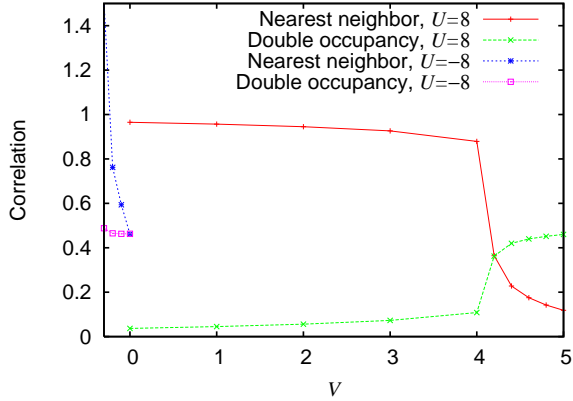}
\caption{(Color online) 
Double occupancy $ \langle n_{i,\uparrow} n_{i,\downarrow} \rangle $ and nearest-neighbor charge-density correlation $ \langle n_i n_{i+1} \rangle $ as a function of $ V $ in ground states with $ U=8 $ and $ -8 $.
\label{fig:12x1_Upm8Vx_donn}}
\end{figure}

Before the analysis of photoinduced states, we show the $ V $ dependence of the double occupancy $ \langle n_{i,\uparrow} n_{i,\downarrow} \rangle $ and the nearest-neighbor charge-density correlation $ \langle n_i n_{i+1} \rangle $ in the ground states with $ U=8 $ and $ -8 $ in Fig.~\ref{fig:12x1_Upm8Vx_donn}.
When the interactions are repulsive, the spin-density-wave (SDW) correlation is dominant for $ V < U/2 + \delta $, and the charge-density-wave (CDW) correlation is dominant for $ V > U/2 + \delta $. Here, $ \delta $ is a small positive quantity originating from spin fluctuations that approaches zero in the $ t_0 \rightarrow 0 $ limit. When the SDW correlation is dominant, $ \langle n_{i,\uparrow} n_{i,\downarrow} \rangle $ is small and $ \langle n_i n_{i+1} \rangle $ is almost 1. When the CDW correlation is dominant, $ \langle n_i n_{i+1} \rangle $ is small and $ \langle n_{i,\uparrow} n_{i,\downarrow} \rangle $ is about 0.5. When the interactions are attractive, phase separation is anticipated. In this case, $ \langle n_{i,\uparrow} n_{i,\downarrow} \rangle $ is about 0.5, which is the average of  $ \langle n_{i,\uparrow} n_{i,\downarrow} \rangle \simeq 1 $ on the high-charge-density sites and $ \langle n_{i,\uparrow} n_{i,\downarrow} \rangle \simeq 0 $ on the low-charge-density sites. Meanwhile, $ \langle n_i n_{i+1} \rangle $ is about 1.67, which is the average of $ \langle n_i n_{i+1} \rangle \simeq 4 $ on the bonds between high-charge-density sites and $ \langle n_i n_{i+1} \rangle \simeq 0 $ on the other bonds. In Fig.~\ref{fig:12x1_Upm8Vx_donn}, we show the results for $ V \geq -0.3 $ because the states for $ V \leq -0.4 $ are numerically unstable. 

\begin{figure}
\includegraphics[height=11.2cm]{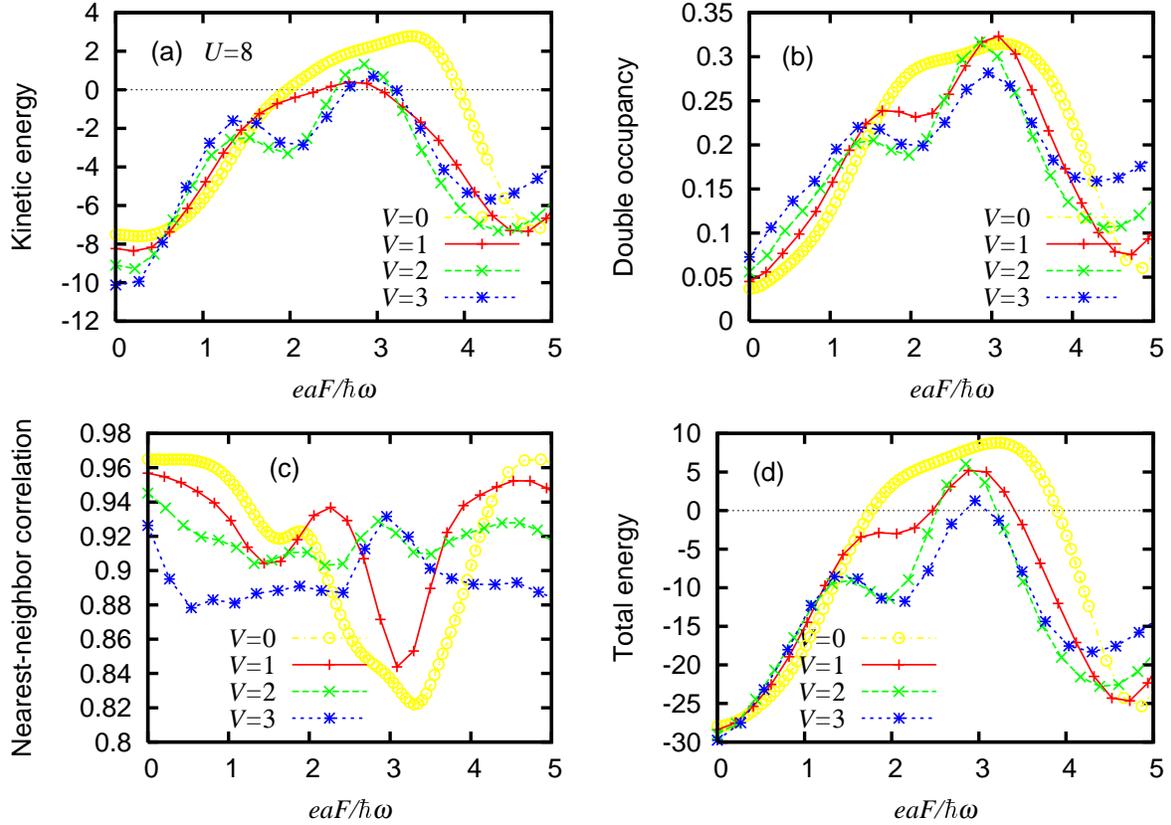}
\caption{(Color online) 
(a) Time-averaged kinetic energy, (b) time-averaged double occupancy $ \langle \langle n_{i,\uparrow} n_{i,\downarrow} \rangle \rangle $, (c) time-averaged nearest-neighbor charge-density correlation $ \langle \langle n_i n_{i+1} \rangle \rangle $, and (d) total energy as functions of $ eaF/\hbar\omega $ for different values of $ V $. The other parameters are $ U=8 $, $ \omega=4.93 $ for $ V=0 $, $ \omega=4.86 $ for $ V=1 $, $ \omega=4.56 $ for $ V=2 $, and $ \omega=3.72 $ for $ V=3 $.
\label{fig:12x1_U8Vx_fx_0to5}}
\end{figure}

Time-averaged quantities for $ U $=8 with different values of $ V $ are shown in Fig.~\ref{fig:12x1_U8Vx_fx_0to5} as functions of $ eaF/\hbar\omega $. 
In all the cases shown here, there is a region around $ eaF/\hbar\omega $=3 for positive time-averaged kinetic energies, although this region is narrowed with increasing $ V $ [Fig.~\ref{fig:12x1_U8Vx_fx_0to5}(a)]. In nearly the same $ eaF/\hbar\omega $ region, the time-averaged double occupancy $ \langle \langle n_{i,\uparrow} n_{i,\downarrow} \rangle \rangle $ becomes larger than 0.25 [Fig.~\ref{fig:12x1_U8Vx_fx_0to5}(b)]: the effective on-site interaction is attractive. On the other hand, where the quantities shown in Figs.~\ref{fig:12x1_U8Vx_fx_0to5}(a) and \ref{fig:12x1_U8Vx_fx_0to5}(b) are at local minima around $ eaF/\hbar\omega $=4.5, the time-averaged nearest-neighbor charge-density correlation $ \langle \langle n_i n_{i+1} \rangle \rangle $ is at a local maximum [Fig.~\ref{fig:12x1_U8Vx_fx_0to5}(c)], and the total energy is at a local minimum [Fig.~\ref{fig:12x1_U8Vx_fx_0to5}(d)]. The corresponding time-averaged state approaches the ground state, although the former is not so close to the latter compared with the corresponding time-averaged state in the one-dimensional three-quarter-filled strongly dimerized extended Peierls-Hubbard model, whose total-energy increment is described reasonably well by the zeroth-order Bessel function multiplied by the sine function\cite{yonemitsu_jpsj15} as in the two-level system.\cite{nishioka_jpsj14} 

Where the quantities shown in Figs.~\ref{fig:12x1_U8Vx_fx_0to5}(a) and \ref{fig:12x1_U8Vx_fx_0to5}(b) are at the maxima around $ eaF/\hbar\omega $=3, $ \langle \langle n_i n_{i+1} \rangle \rangle $ is at a local maximum (minimum) for $ V $=2 and 3 ($ V $=0 and 1) [Fig.~\ref{fig:12x1_U8Vx_fx_0to5}(c)], and the total energy is positive and at the maximum [Fig.~\ref{fig:12x1_U8Vx_fx_0to5}(d)]. The state realized here is a negative-temperature state. The fact that, for $ V $=2 and 3, both $ \langle \langle n_{i,\uparrow} n_{i,\downarrow} \rangle \rangle $ and $ \langle \langle n_i n_{i+1} \rangle \rangle $ are at local maxima in the negative-temperature state indicates that phase separation is realized here. To see how roughly the negative-temperature state is interpreted as the ground state with the inverted interactions, we compare the correlations in the ground states for $ U $=8 and $ V>0 $ and for $ U $=$-$8 and $ V<0 $ (Fig.~\ref{fig:12x1_Upm8Vx_donn}). For small and positive $ V $ ($ V $=0.1 and 0.2), $ \langle n_i n_{i+1} \rangle $ is decreased by the inversion of the interactions. This is in contrast to the cases of $ V \geq 0.3 $, where both $ \langle n_{i,\uparrow} n_{i,\downarrow} \rangle $ and $ \langle n_i n_{i+1} \rangle $ are increased by the inversion. The local minimum of $ \langle \langle n_i n_{i+1} \rangle \rangle $ around $ eaF/\hbar\omega $=3 for $ V $=1 [Fig.~\ref{fig:12x1_U8Vx_fx_0to5}(c)] should be caused by the fact that $ V $ is small and the inversion of the interactions is incomplete. As a matter of fact, $ \langle \langle n_i n_{i+1} \rangle \rangle $ at the local maximum is smaller (only slightly larger) than $ \langle n_i n_{i+1} \rangle $ in the ground state for $ V $=2 ($ V $=3). Thus, the overall behaviors of the correlations in the negative-temperature states are consistent with those in the ground states with incompletely inverted interactions. Compared with the inverted on-site interaction, which makes the double occupancy larger than 0.25, the inverted intersite interaction is less effective. Note that, in the ground state, the value of $ \mid V \mid $ required for phase separation ($ U $, $ V<0 $) is generally much smaller than that for the CDW ($ U $, $ V>0 $). In the thermodynamic limit and in the $ t_0 \rightarrow 0 $ limit, the former approaches zero but the latter is proportional to $ U $ with a dimension-dependent coefficient. From this viewpoint,  even if the inversion of the intersite interaction is incomplete, photoinduced phase separation is feasible. 

\subsection{Two-dimensional case}

\begin{figure}
\includegraphics[height=5.6cm]{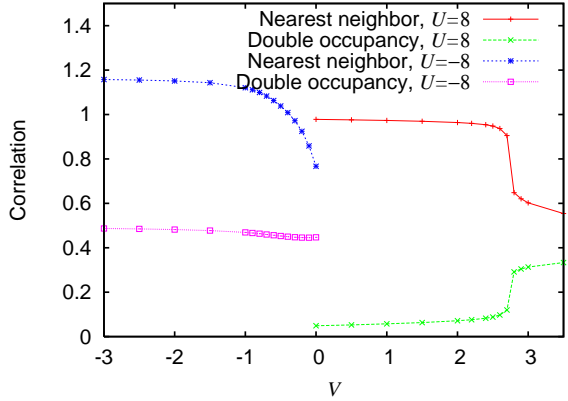}
\caption{(Color online) 
Double occupancy $ \langle n_{i,\uparrow} n_{i,\downarrow} \rangle $ and average nearest-neighbor charge-density correlation $ \frac{1}{2} \langle n_i n_{i+\hat{x}}+ n_i n_{i+\hat{y}} \rangle $ as a function of $ V $ in ground states with $ U=8 $ and $ -8 $.
\label{fig:4x3_Upm8Vx_donn}}
\end{figure}

First, we show the $ V $ dependence of the double occupancy $ \langle n_{i,\uparrow} n_{i,\downarrow} \rangle $ and the average nearest-neighbor charge-density correlation $ \frac{1}{2} \langle n_i n_{i+\hat{x}}+ n_i n_{i+\hat{y}} \rangle $, where $\hat{x}$ and $\hat{y}$ are the unit vectors along the $ x $- and $ y $-axes, respectively, in the ground states with $ U=8 $ and $ -8 $ in Fig.~\ref{fig:4x3_Upm8Vx_donn}.
When the interactions are repulsive, the SDW correlation is dominant for $ V < U/4 + \delta $, and the CDW correlation is dominant for $ V > U/4 + \delta $. Here, $ \delta $ is not so small because the present $ 4 \times 3 $ lattice is incommensurate with the CDW with the wave vector of ($\pi$,$\pi$). In the thermodynamic limit and in the $ t_0 \rightarrow 0 $ limit, $ \delta $ approaches zero. Compared with the one-dimensional case, the numerical instability for attractive interactions is greatly suppressed. In the case of phase separation, $ \langle n_{i,\uparrow} n_{i,\downarrow} \rangle $ is about 0.5 and $ \frac{1}{2} \langle n_i n_{i+\hat{x}}+ n_i n_{i+\hat{y}} \rangle $ is about 1.17, which is the average of $ \frac{1}{2} \langle n_i n_{i+\hat{x}}+ n_i n_{i+\hat{y}} \rangle \simeq 4 $ on the bonds between high-charge-density sites and $ \frac{1}{2} \langle n_i n_{i+\hat{x}}+ n_i n_{i+\hat{y}} \rangle \simeq 0 $ on the other bonds. Note that the weight of the boundary bonds between the high-charge-density sites and the low-charge-density sites is much larger than that in the one-dimensional case, so that the nearest-neighbor charge-density correlation here is substantially smaller than that in the one-dimensional case. In other words, the boundary degrees of freedom are much larger in the $ 4 \times 3 $ lattice than in the 12-site chain, so that the tendency toward phase separation is much weaker in the $ 4 \times 3 $ lattice. 

\begin{figure}
\includegraphics[height=11.2cm]{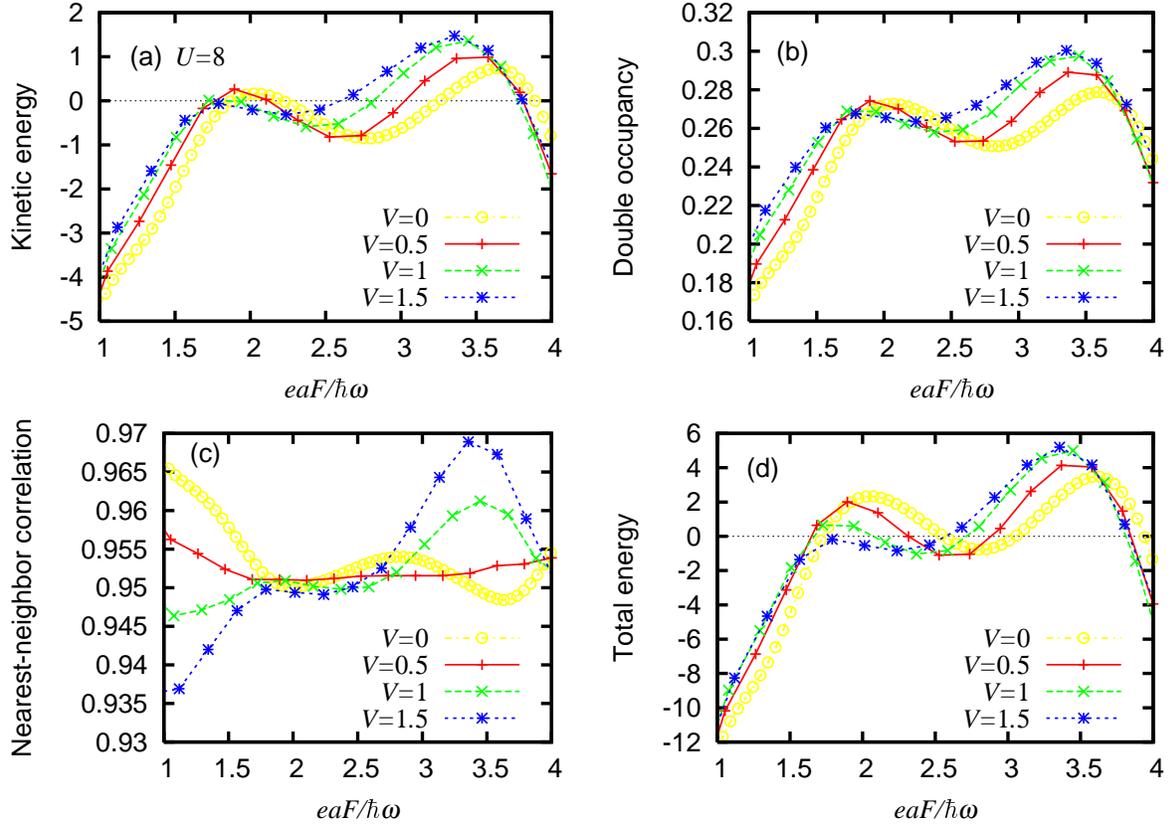}
\caption{(Color online) 
(a) Time-averaged kinetic energy, (b) time-averaged double occupancy $ \langle \langle n_{i,\uparrow} n_{i,\downarrow} \rangle \rangle $, (c) time-averaged nearest-neighbor charge-density correlation $ \frac{1}{2} \langle \langle n_i n_{i+\hat{x}}+ n_i n_{i+\hat{y}} \rangle \rangle $, and (d) total energy as functions of $ eaF/\hbar\omega $ for different values of $ V $. The other parameters are $ U=8 $, $ \omega=4.82 $ for $ V=0 $, $ \omega=4.75 $ for $ V=0.5 $, $ \omega=4.64 $ for $ V=1 $, and $ \omega=4.47 $ for $ V=1.5 $.
\label{fig:4x3_U8Vx1p5_fx_1to4}}
\end{figure}

Time-averaged quantities for $ U $=8 with different values of $ V $ are shown in Fig.~\ref{fig:4x3_U8Vx1p5_fx_1to4} as functions of $ eaF/\hbar\omega $. 
Compared with the $ V/2 $ dependence of the corresponding quantities in the one-dimensional case, the present $ V $ dependence is generally small, which is consistent with the fact that the tendency toward phase separation is weaker in the two-dimensional case. Then, we show the region of $ 1< eaF/\hbar\omega < 4$ to enlarge the view around the local maxima of the total energy. The $ eaF/\hbar\omega $ region for positive time-averaged kinetic energies is shifted to the smaller $ eaF/\hbar\omega $ side with increasing $ V $ [Fig.~\ref{fig:4x3_U8Vx1p5_fx_1to4} (a)]. In a relatively wide region including the $ eaF/\hbar\omega $ region above, $ \langle \langle n_{i,\uparrow} n_{i,\downarrow} \rangle \rangle $ becomes larger than 0.25 [Fig.~\ref{fig:4x3_U8Vx1p5_fx_1to4}(b)]: the effective on-site interaction is attractive whenever the time-averaged kinetic energy is positive. 

Where the quantities shown in Figs.~\ref{fig:4x3_U8Vx1p5_fx_1to4}(a) and \ref{fig:4x3_U8Vx1p5_fx_1to4}(b) are at the maxima around $ eaF/\hbar\omega $=3.5, $ \frac{1}{2} \langle \langle n_i n_{i+\hat{x}}+ n_i n_{i+\hat{y}} \rangle \rangle $ is at a local maximum (minimum) for $ V $=1 and 1.5 ($ V $=0) [Fig.~\ref{fig:4x3_U8Vx1p5_fx_1to4}(c)], and the total energy is positive and at the maximum [Fig.~\ref{fig:4x3_U8Vx1p5_fx_1to4}(d)]. Negative-temperature states are realized here in a wide $ eaF/\hbar\omega $ region because the $ U $ term assists the total energy to become positive, while the $ V $ term does not assist it at all ($ \frac{1}{2} \langle \langle n_i n_{i+\hat{x}}+ n_i n_{i+\hat{y}} \rangle \rangle $ is always smaller than 1). The fact that, for $ V $=1, 1.5 (shown here), 2, 2.5, and 3 (not shown here), both $ \langle \langle n_{i,\uparrow} n_{i,\downarrow} \rangle \rangle $ and $ \frac{1}{2} \langle \langle n_i n_{i+\hat{x}}+ n_i n_{i+\hat{y}} \rangle \rangle $ are at local maxima in the negative-temperature state indicates that phase separation is realized around $ eaF/\hbar\omega $=3.5 in the two-dimensional case. Then, we compare the correlations in the ground states for $ U $=8 and $ V>0 $ and for $ U $=$-$8 and $ V<0 $ (Fig.~\ref{fig:4x3_Upm8Vx_donn}). For small and positive $ V $ ($ V $=0.1 and 0.2), $ \frac{1}{2} \langle \langle n_i n_{i+\hat{x}}+ n_i n_{i+\hat{y}} \rangle \rangle $ is decreased by the inversion of the interactions. The absence of an apparent local maximum of $ \frac{1}{2} \langle \langle n_i n_{i+\hat{x}}+ n_i n_{i+\hat{y}} \rangle \rangle $ around $ eaF/\hbar\omega $=3.5 for $ V $=0.5 [Fig.~\ref{fig:4x3_U8Vx1p5_fx_1to4}(c)] should again be caused by the fact that $ V $ is small and the inversion of the interactions is incomplete. As a matter of fact, $ \frac{1}{2} \langle \langle n_i n_{i+\hat{x}}+ n_i n_{i+\hat{y}} \rangle \rangle $ at the local maximum is smaller (larger) than $ \frac{1}{2} \langle n_i n_{i+\hat{x}}+ n_i n_{i+\hat{y}} \rangle $ in the ground state for $ V $=1 and 1.5 ($ V $=2, 2.5, and 3) (not shown). Again, the overall behaviors of the correlations in the negative-temperature states are consistent with those in the ground states with incompletely inverted interactions. Even though $ \frac{1}{2} \langle \langle n_i n_{i+\hat{x}}+ n_i n_{i+\hat{y}} \rangle \rangle $ at the local maximum is still smaller than 1 in all cases in the present small system to which the exact diagonalization method is applicable, a strong photoexcitation may lead to phase separation in much larger systems. 

Recall that the formation of a negative-temperature state is generally obstructed by a large rise in the entropy. The photoinduced phase separation is not obstructed by such an effect. However, when we employ attractive interactions ($ U $, $ V<0 $) and strongly excite phase-separating states, no negative-temperature state is realized (not shown). The total energy and the time-averaged kinetic energy remain negative for any $ eaF/\hbar\omega $ under both resonant and nonresonant conditions. The time-averaged double occupancy is always larger than 0.25. These facts indicate that it is difficult to increase the total energy of an initially phase-separating state without a large rise in the entropy. The entropy is easily raised by exciting low-energy collective states corresponding to the fluctuating boundary. 

\subsection{Hartree-Fock approximation for two-dimensional case}

It is difficult to treat larger systems at half filling for long periods of time by the exact diagonalization method. However, in order to judge whether phase separation is really achieved, we need to study the photoinduced dynamics of larger systems anyway. Hereafter, we use the Hartree-Fock approximation for larger systems with a periodic boundary condition on the square lattice and compare the results for different system sizes. The ground state is an SDW state. The central frequency $ \omega $ is chosen to be nearly resonant with the optical gap in the time-dependent Hartree-Fock approximation. In expanding the exponential evolution operator, a time slice of $ dt =10^{-3} $ is used. The time averaging is performed using Eq.~(\ref{eq:time_average}) with $ t_s=100 T $ and $ t_w=50 T $. At first, in order to compare results obtained by these methods, we show the same quantities as in Fig.~\ref{fig:4x3_U8Vx1p5_fx_1to4} as functions of $ eaF/\hbar\omega $. Here, the kinetic and total energies are divided by the number of sites for the comparison of results with different system sizes. The data shown here are obtained for the 10$ \times $10 lattice, and they coincide with the results for the 14$ \times $14 and 16$ \times $16 lattices within the symbol sizes. Although the value of $ eaF/\hbar\omega $ at which a local maximum/minimum of each quantity appears deviates from the corresponding one obtained by the exact diagonalization method, these results obtained with the different methods show qualitatively similar $ eaF/\hbar\omega $ and $ V $ dependences. 

\begin{figure}
\includegraphics[height=11.2cm]{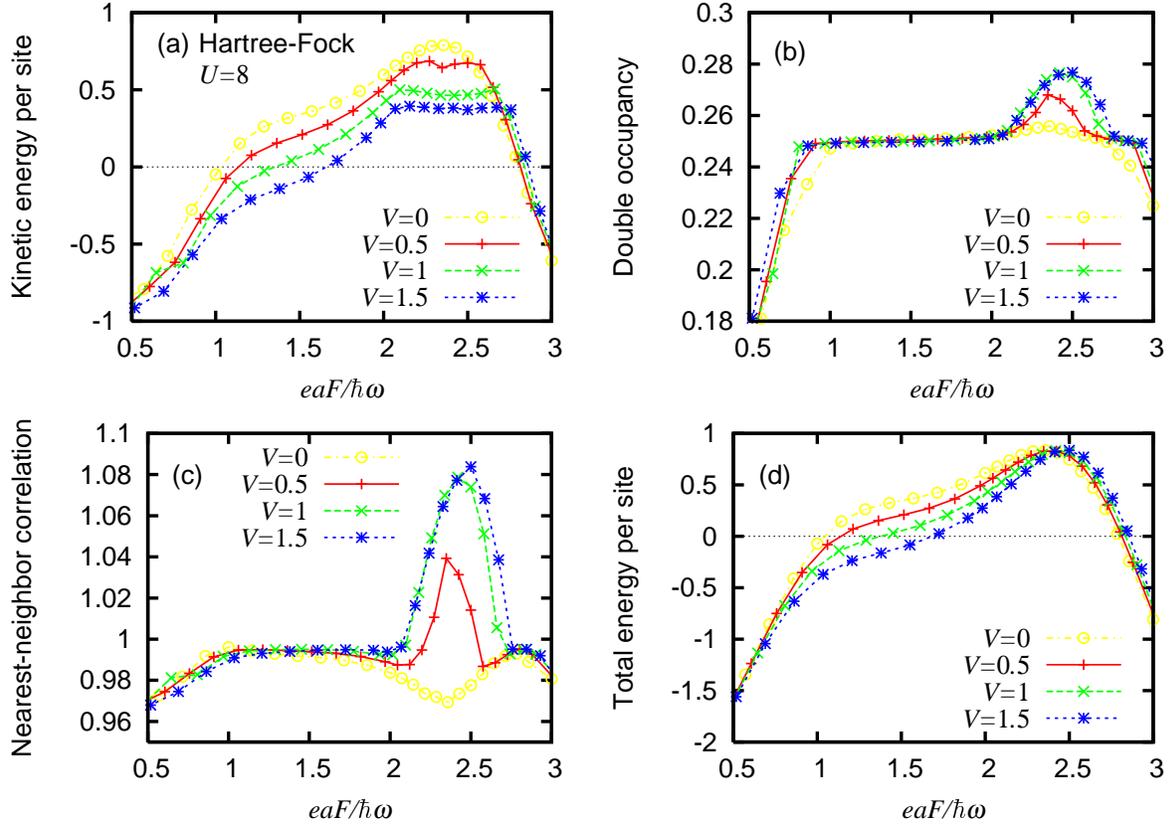}
\caption{(Color online)
(a) Time-averaged kinetic energy per site, (b) time-averaged double occupancy $ \langle \langle n_{i,\uparrow} n_{i,\downarrow} \rangle \rangle $, (c) time-averaged nearest-neighbor charge-density correlation $ \frac{1}{2} \langle \langle n_i n_{i+\hat{x}}+ n_i n_{i+\hat{y}} \rangle \rangle $, and (d) total energy per site as functions of $ eaF/\hbar\omega $ obtained in Hartree-Fock approximation for different values of $ V $. The other parameters are $ U=8 $, $ \omega=7.0 $ for $ V=0 $, $ \omega=6.6 $ for $ V=0.5 $, $ \omega=6.2 $ for $ V=1 $, and $ \omega=5.8 $ for $ V=1.5 $.
\label{fig:10x10_U8Vx1p5_fx_0p5to3}}
\end{figure}

In a region around $ eaF/\hbar\omega $=2.4, the time-averaged kinetic energy and the total energy are positive [Figs.~\ref{fig:10x10_U8Vx1p5_fx_0p5to3}(a) and \ref{fig:10x10_U8Vx1p5_fx_0p5to3}(d)]. 
However, their $ V $ dependence is opposite to that obtained by the exact diagonalization method, although the $ V $ dependence itself is rather weak. The total energy shows a local maximum near $ eaF/\hbar\omega $=2.4, where $ \langle \langle n_{i,\uparrow} n_{i,\downarrow} \rangle \rangle $ becomes larger than 0.25 [Fig.~\ref{fig:10x10_U8Vx1p5_fx_0p5to3}(b)]. The effective on-site attraction increases with $ V $. Around $ eaF/\hbar\omega $=2.4, $ \frac{1}{2} \langle \langle n_i n_{i+\hat{x}}+ n_i n_{i+\hat{y}} \rangle \rangle $ is at a local maximum (minimum) for $ V $=0.5, 1, and 1.5 ($ V $=0) [Fig.~\ref{fig:10x10_U8Vx1p5_fx_0p5to3}(c)]. Now, $ \frac{1}{2} \langle \langle n_i n_{i+\hat{x}}+ n_i n_{i+\hat{y}} \rangle \rangle $ exceeds 1, but its maximum value is slightly over 1. The behavior around $ eaF/\hbar\omega $=2.4 is consistent with the appearance of phase separation, but $ \frac{1}{2} \langle \langle n_i n_{i+\hat{x}}+ n_i n_{i+\hat{y}} \rangle \rangle $ is considerably smaller than that in the phase-separating ground state with the inverted on-site and intersite interactions. This fact will be clarified below when the space and time dependences are discussed. The maximum value of $ \frac{1}{2} \langle \langle n_i n_{i+\hat{x}}+ n_i n_{i+\hat{y}} \rangle \rangle $ is insensitive to the system size, at least up to the 16$\times$16 lattice. As far as the quantities shown here are concerned, they should already be close to those in the thermodynamic limit. However, the space and time dependences can be sensitive to the system size. 

\begin{figure}
\includegraphics[height=11.2cm]{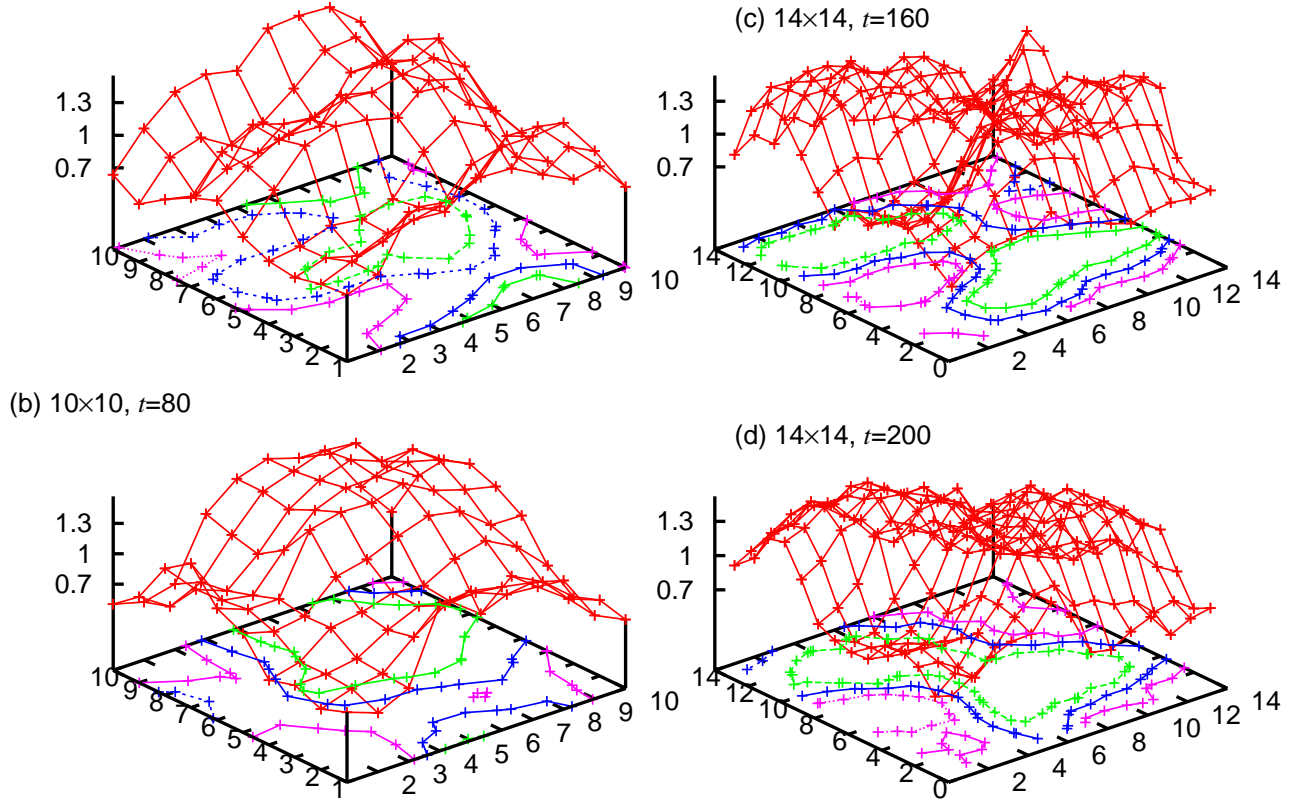}
\caption{(Color online)
Snapshots of charge density $ \langle n_{i} \rangle $ at (a) $t$=60 on 10$\times$10 lattice, (b) $t$=80 on 10$\times$10 lattice, (c) $t$=160 on 14$\times$14 lattice, and (d) $t$=200 on 14$\times$14 lattice obtained by Hartree-Fock approximation for $ U=8 $, $ V=1 $, $ \omega=6.2 $, and $ eaF/\hbar\omega=2.4 $.
\label{fig:100_196_U8V1w6p2f15_4ts}}
\end{figure}

Snapshots of the charge-density distribution $ \langle n_{i} \rangle $ are shown in Fig.~\ref{fig:100_196_U8V1w6p2f15_4ts} at different times and on the 10$\times$10 and 14$\times$14 lattices. 
The contour lines for $ \langle n_{i} \rangle $=0.7, 1.0, and 1.3 are shown on their bases. The initial condition before the photoexcitation is spatially uniform, so that $ \langle n_{i} \rangle $=1 at $ t $=0. However, the translational symmetry becomes spontaneously broken after the photoexcitation. Then, the charge-density distribution $ \langle n_{i} \rangle $ fluctuates spatially and temporally. At first, a large number of small high-density regions and low-density regions appear. As time proceeds, some of them merge into larger high-density regions and low-density regions. On the 10$\times$10 lattice, there are two high-density regions at $ t $=60 [Fig.~\ref{fig:100_196_U8V1w6p2f15_4ts}(a)], and they have already merged into one at $ t $=80 [Fig.~\ref{fig:100_196_U8V1w6p2f15_4ts}(b)]. On the 14$\times$14 lattice, there are two high-density regions at $ t $=160 [Fig.~\ref{fig:100_196_U8V1w6p2f15_4ts}(c)], and they are about to merge into one at $ t $=200 [Fig.~\ref{fig:100_196_U8V1w6p2f15_4ts}(d)]. For the 16$\times$16 lattice, the time when the high-density regions finally merge into one is later than that for the 14$\times$14 lattice. 

Such a size dependence of domain dynamics is actually reflected in a small but significant size dependence of $ \frac{1}{2} \langle \langle n_i n_{i+\hat{x}}+ n_i n_{i+\hat{y}} \rangle \rangle $. For $ U $=8, $ V $=1, and $ eaF/\hbar\omega $=2.4, where $ \frac{1}{2} \langle \langle n_i n_{i+\hat{x}}+ n_i n_{i+\hat{y}} \rangle \rangle $ shows a maximum, $ \frac{1}{2} \langle \langle n_i n_{i+\hat{x}}+ n_i n_{i+\hat{y}} \rangle \rangle $=1.079 on the 10$\times$10 lattice where the high-density regions have already merged into one before $ t_s $ (the time averaging), $ \frac{1}{2} \langle \langle n_i n_{i+\hat{x}}+ n_i n_{i+\hat{y}} \rangle \rangle $=1.070 on the 14$\times$14 lattice where the high-density regions have not yet merged into one at $ t_s+t_w $, and $ \frac{1}{2} \langle \langle n_i n_{i+\hat{x}}+ n_i n_{i+\hat{y}} \rangle \rangle $=1.074 on the 16$\times$16 lattice where the situation is similar to that on the 14$\times$14 lattice. On the 14$\times$14 lattice, the existence of more than one domain makes the boundary contribution larger and the above correlation smaller than those on the 10$\times$10 lattice. On the 16$\times$16 lattice, the larger system size makes the boundary contribution smaller and the above correlation larger than those on the 14$\times$14 lattice. The size dependence of the time-averaged correlation above thus originates from the timing of the merging and the relative weight of the boundary degrees of freedom. In any case, the time-averaged correlation above is almost saturated for these system sizes and the size dependence is small. 

The charge density in the high-density regions is not as high as $ \langle n_{i} \rangle \simeq 2 $, and that in the low-density regions is not as low as $ \langle n_{i} \rangle \simeq 0 $. This is regarded as being due to a rise in the entropy. The rise in the entropy is not so large as to obstruct the formation of a negative-temperature state, but it is significant, so that the charge disproportionation is small. This is the reason why the quantities shown in Fig.~\ref{fig:10x10_U8Vx1p5_fx_0p5to3} are insensitive to the system size and the reason why the maximum value of $ \frac{1}{2} \langle \langle n_i n_{i+\hat{x}}+ n_i n_{i+\hat{y}} \rangle \rangle $ is slightly over 1. Even though the charge disproportionation is small, Fig.~\ref{fig:100_196_U8V1w6p2f15_4ts} shows a nonequilibrium phase-separating state, which is quantitatively different from the phase-separating ground state with the completely inverted on-site and intersite interactions. The inversion of the intersite interaction is incomplete, namely, the effective intersite interaction becomes negative, but its magnitude is smaller than the initial value. Therefore, the previous results obtained by the exact diagonalization method for a small system are consistent with the present Hartree-Fock results and the anticipation of photoinduced phase separation in sufficiently large systems. 

\section{Conclusions and Discussion}

Many-electron dynamics in the one- and two-dimensional half-filled extended Hubbard models after the application of a symmetric monocycle electric-field pulse are calculated using the exact diagonalization method for small systems and the Hartree-Fock approximation for large systems. The appearance of negative-temperature states is demonstrated by their total energies higher than that in an equilibrium state at infinite temperature. They are characterized by inverted on-site and nearest-neighbor interactions, which are roughly estimated from the field-amplitude dependence of time-averaged correlation functions. The important point is that, even by applying symmetric monocycle pulses and even in models on regular lattices without dimerization, negative-temperature states are realized in a wide region spanned by the field amplitude and the photoexcitation energy. Under nearly the same conditions, the time-averaged double occupancy is larger than that of the noninteracting equilibrium state, and thus the on-site interaction is effectively inverted. In addition, if the nearest-neighbor repulsion is not very small, the time-averaged nearest-neighbor charge-density correlation shows a local maximum as a function of the field amplitude. Although the inversion of the interactions is incomplete, the intersite interaction is also inverted. These field-amplitude and intersite-interaction dependences of the time-averaged correlation functions are consistent with the anticipated photoinduced phase separation. 

The difficulty of forming a negative-temperature state by applying a symmetric monocycle pulse in the dynamical mean-field theory for the half-filled Hubbard model\cite{tsuji_prb12} may be caused by the fact that Umklapp scattering is included indirectly through the self-energy in the theory, where the momentum space is basically irrelevant. The fact that the kinetic energy becomes positive and consequently the momentum distribution does not return to the initial one after intense irradiation of a symmetric monocycle pulse suggests the importance of the Umklapp process. Thus, it would be desirable to treat the momentum space or the real space in a direct manner. The present approach is expected to be suitable for the discussion of negative-temperature states. 
Photoinduced superconductivity is proposed using the Hubbard model.\cite{tsuji_prl11} Real substances always have intersite repulsions, so that photoinduced phase separation can be realized once the intersite interactions are effectively inverted: not only two electrons but many electrons attract each other. As to phase separation, a spatially extended region must be treated to account for the gradual space dependence of the charge density. In this context, it would be better to supplement the exact diagonalization study with Hartree-Fock studies, as in this paper. Generally, a superconducting state with a high transition temperature is expected to be realized near some electronic instability. Phase separation has been discussed extensively in this context. Namely, in the vicinity of a phase-separating state, superconductivity with an electronic mechanism has been anticipated.\cite{ogata_prl91,hellberg_prl97,voit_prb92,penc_prb94,kuroki_prb94,clay_prb99,aligia_prb00,su_prb04,vandongen_prl95} As the possibility of photoinduced phase separation is enhanced, that of photoinduced superconductivity should also be enhanced. 

In our numerical calculations, as the system size increases, it requires a longer time for high-density or low-density regions to merge into one. This is due to the initial condition in which the charge density is spatially uniform. In real substances at finite temperatures, electron-phonon interactions exist and thermal lattice fluctuations should exert a random force on electrons, so that high-density or low-density regions should easily appear, be activated to move around, and encounter similar-density regions that they merge with. Unless the system is biased externally, however, such regions can appear anywhere in the sample. We expect that an external bias such as a DC electric field would assist macroscopic phase separation with an anomalous dielectric property. Although the electron configuration in the ground state is determined so that the total energy is minimized, that in the photoinduced state with the largest total energy is determined in the opposite manner: the instability is a key because the realized state is the most unstable state among the photoinduced states. We expect that the negative-temperature state in real substances would decay rapidly through every channel, interacting with phonons, photons, and electrons in the environment. The lifetime of the negative-temperature state would be short especially when the system has a complex structure and interacts with many degrees of freedom. 

\begin{acknowledgment}

This work was supported by a Grant-in-Aid for Scientific Research (C) (Grant No. 23540426) from the Ministry of Education, Culture, Sports, Science and Technology of Japan. 

\end{acknowledgment}

\appendix
\section{Negative-Temperature States Induced by Half-Cycle Pulse}

Here only, we consider the half-cycle cosine pulse used in Ref.~\citen{tsuji_prb12}, which is in the present formulation written as 
\begin{equation}
\mbox{\boldmath $A$} (t) = -\frac{\hbar c}{ea} (A,A) 
\left[ \frac{t}{\tau} -\frac{1}{2\pi} \sin\left(\frac{2\pi t}{\tau}\right) \right] 
\theta (t) \theta ( \tau-t )
\;, \label{eq:half_cycle_pulse}
\end{equation}
with $ \tau $ being the pulse width. It is applied to the half-filled Hubbard model ($ V = 0 $) on the 4$\times$3 lattice. We use the notation $ A $, as in Ref.~\citen{tsuji_prb12}, which is the magnitude of the Peierls phase shift for the electron transfer in any direction. 

\begin{figure}
\includegraphics[height=5.6cm]{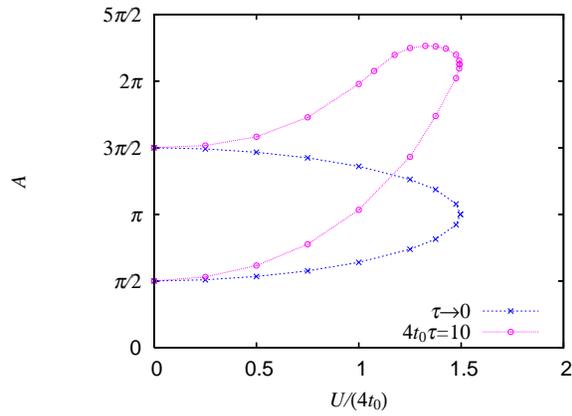}
\caption{(Color online)
Region where total energy is positive, around point ($U$, $A$)=(0, $\pi$) in plane spanned by $U/(4t_0)$ and $A$, for Hubbard model driven by half-cycle cosine pulse with width $\tau \rightarrow 0$ or $ 4t_0 \tau = 10 $. 
\label{fig:4x3_halfccl}}
\end{figure}

The energy dependence of the density of states is different from that used in Ref.~\citen{tsuji_prb12}. Then, here, we take the half-bandwidth in the noninteracting case, $ 4t_0 $, as a unit of energy. The dimensionless repulsion strength is $ U/(4t_0) $ and the dimensionless pulse width is $ 4t_0 \tau $. The region where the total energy is positive after the photoexcitation, i.e. where negative-temperature states appear, is surrounded by the dashed line for the delta-function pulse ($\tau \rightarrow 0$) and by the dotted line for the pulse with width $ 4t_0 \tau = 10 $ in Fig.~\ref{fig:4x3_halfccl}. 
For the delta-function pulse, this region is symmetric with respect to the $ A=\pi $ line. For the pulse with finite width, this region is shifted to the larger $ A $ side as $ U/(4t_0) $ increases. These results are consistent with those in the dynamical mean-field theory.\cite{tsuji_prb12} A negative-temperature state is formed by the Peierls phase shift as well in the present system. 

% Create the reference section using BibTeX:
\bibliography{67116}

\end{document}